\documentclass[11pt,twoside]{article}
\usepackage{asp2004}
\usepackage{psfig}
\usepackage{epsf}
\usepackage{graphics}
\usepackage{lscape}
\newcommand{\HI}{H{\,\small I}}
\def\edcomment#1{\iffalse\marginpar{\raggedright\sl#1\/}\else\relax\fi}
\reversemarginpar

\begin{document}
\title{Fast Outflow of Neutral and Ionized Gas from the Radio Galaxy 3C 293}
\author{B.H.C. Emonts}
\affil{Kapteyn Astronomical Institute, Postbus 800, 9700 AV Groningen, The Netherlands}
\author{R. Morganti, T.A. Oosterloo}
\affil{Netherlands Foundation for Research in Astronomy, Postbus 2, NL-7990 AA Dwingeloo, The Netherlands}
\author{C.N. Tadhunter}
\affil{Department of Physics and Astronomy, University of Sheffield, Hicks Building, Hounsfield Road, Sheffield S7 3RH, UK}
\author{J.M. van der Hulst}
\affil{Kapteyn Astronomical Institute, Postbus 800, 9700 AV Groningen, The Netherlands}

\begin{abstract}
We detect a fast outflow of neutral and ionized gas with velocities up to $\sim$1000 km/s from the central region of radio galaxy 3C 293. With optical spectroscopy we locate the bulk of the ionized gas outflow at the position of a bright radio hot-spot in the inner radio jet, about 1 kpc east of the nucleus. Given the presence of large amounts of cold gas and the distorted morphology of the radio jet in this region, we argue that the ISM is pushed out by a severe interaction with the radio plasma. The similarity of the outflow of \HI\ with the ionized gas outflow that we see at the position of the radio hot-spot suggests that despite the high energies involved in the jet-ISM interaction, part of the gas stays, or becomes again, neutral. In this paper we also present the detection of \HI\ emission in three nearby companions of 3C 293. 
\end{abstract}
\thispagestyle{plain}

\section{Introduction}

AGN and starburst activity are important physical processes that influence the ISM in the central regions of active galaxies. Besides their strong radiation-field that ionizes the gas, jet-cloud interactions, as well as AGN and starburst driven winds, shock the gas and may create (in some cases massive) outflows. Feedback mechanisms of this activity can be important for the evolution of the host galaxy. The AGN activity may e.g. regulate the correlation between the mass of the central black hole and the galaxies bulge properties, while supernova explosions and galactic winds are processes that regulate star formation in galaxies.

Due to the complex gas dynamics and ionization-processes in the active central regions, the study of interactions between ISM and radio/starburst activity is often quite complicated. Nearby radio galaxies provide excellent cases to study these physical processes in great detail, which is important for understanding the evolution of high-{\it z} radio galaxies and quasars. In a number of nearby radio galaxies unambiguous evidence exists for outflows in the ionized gas, ascribed to the presence of a central radio source \citep[e.g.][]{hol03,tad01,vil99}. In the Seyfert galaxy Mrk 3, \citet{cap99} have studied the impact of a radio jet on the ionized ISM in great detail. However, to get a complete picture on the outflow-related processes it is essential to look at the different phases of the ISM. For example, in addition to the fast ionized gas outflow in PKS 1345+12 \citep{hol03}, \citet{mor04} have observed a fast outflow of neutral hydrogen gas detected in absorption against the radio continuum. With similar \HI\ absorption measurements \citet{oma02} detected a \HI\ outflow in the Seyfert galaxy Mrk 1. And in Seyfert IC 5063 \citet{oos00} detected an outflow of \HI\ at the location of the extended radio lobe \citep[very similar to an outflow of ionized gas;][]{mor04b}. In a recent paper \citep[][hereafter Paper 1]{mor03} we discovered a fast outflow of neutral hydrogen  (with velocities up to $\sim$1000 km/s!) in the nearby radio galaxy 3C 293. In this paper we further analyze this outflow, both in the neutral as well as the ionized gas.

\section{3C 293}
\label{sec:3c293}

The radio continuum image of  Figure \ref{fig:ionized} (left) shows that 3C 293 has two extended radio lobes as well as a Steep Spectrum Core (consisting of a flat spectrum nucleus and two inner radio jets), possibly indicating re-started activity \citep{aku96}. The host galaxy of 3C 293 has a complex, disk-like optical morphology that extends toward an apparent companion \citep{eva99,bre84}. \citet{tad05} revealed young stellar populations throughout the galaxy (the age of this population in the nuclear region is 1.0 - 2.5 Gyr). The central few kpc of the galaxy contains extensive filamentary dust lanes and large amounts of cold gas \citep{bes04,bes02,eva99,has85,baa81}. The presence of a dense ISM, a young stellar population and a distorted radio morphology makes 3C 293 an excellent case for a study of the interplay between ISM, star-formation history and AGN activity. Throughout this paper we use H$_{0}$ = 71 km s$^{-1}$ Mpc$^{-1}$, putting 3C 293 at a distance of 189 Mpc and 1 arcsec $\sim$ 0.92 kpc.

\section{Ionized Gas Outflow}
\label{sec:ionized}

\begin{figure}
\plotone{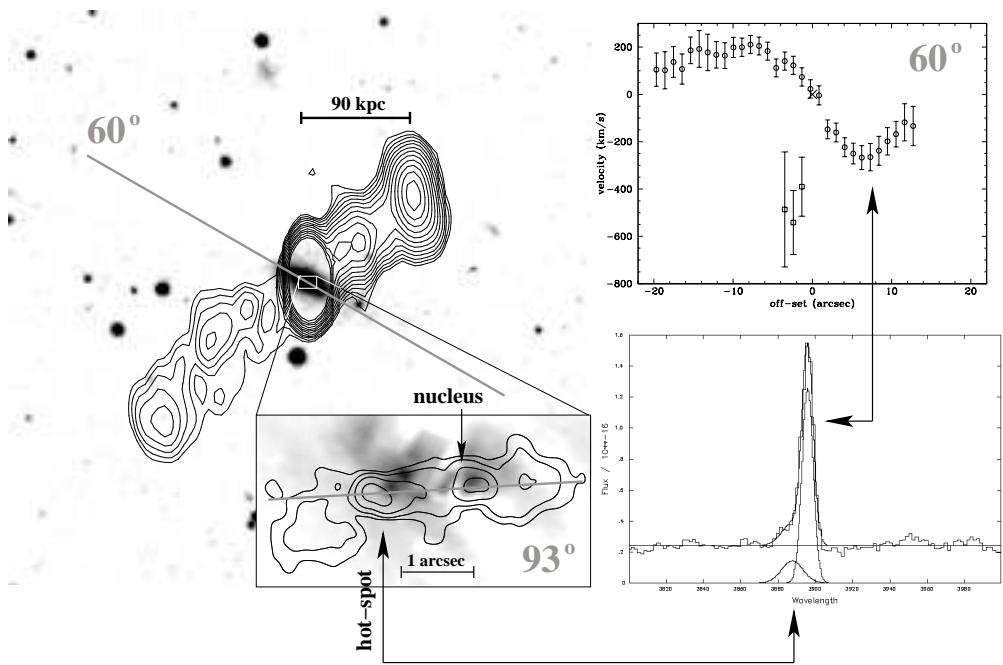}
\caption{{\em Left:} Radio continuum image (from our 10MHz WSRT data, Section \ref{sec:environment}) overlaid on optical DSS image of 3C 293 (contour levels: 4.5, 6.5, 9.0, 13, 18, 25, 35, 50, 70, 95, 130, 170 mJy/beam). The zoom-in on the central few kpc \citep[from][]{bes02} shows a MERLIN image of the radio continuum (contour levels: 9, 36, 144, 630 mJy/beam) overlaid on an optical HST image (the darkest gray-scales represent regions with highest intensity). {\em Bottom right:} Fit to the [OII] line at the position of the radio hot-spot. Visible are both the narrow and the broad component. {\em Top right:} Position-velocity diagram of the [OII] line along the major axis of the galaxy (p.a. 60$^{\circ}$). The circles trace the narrow component; the squares the broad, blueshifted component. The cross marks the systemic velocity (v$_{sys}$=13,450 km/s; see \citet{emo05} for more details).}
\label{fig:ionized}
\end{figure}

We detect a fast outflow of ionized gas at the position of a bright radio hot-spot in the inner radio jet of 3C 293. Evidence comes from a detailed emission-line analysis of optical long-slit spectra that were taken with ISIS on the WHT along the major axis of the host galaxy as well as along the inner radio axis. This is shown in Figure \ref{fig:ionized}.\ [OII]$_{\lambda 3727}$ shows a strong, narrow emission-line component all across 3C 293. This component represents gas with kinematics characteristic for normal rotation along the direction of the optical disk (p.a. 60$^{\circ}$) with a total extent of $\sim$30 kpc, i.e. larger than what has been seen before by \citet{bre84}. When fitting the line-profile in the region of the bright eastern radio hot-spot, $\sim$1 kpc east of the nucleus, we need to include an additional broad emission-line component to get a good fit (see Fig.\ref{fig:ionized}). For this broad component we derive a FWHM of 986 $\pm$ 146 km/s and a blueshift w.r.t. the narrow component of 636 $\pm$ 125 km/s. Together with the distorted morphology of the radio plasma in the hot-spot region \citep{bes04} and the presence of a dense ISM in the inner few kpc of 3C 293, this suggests that the broad component represents an outflow of ionized gas that is driven by interaction with the radio plasma.
\begin{figure}[ht]
\plotone{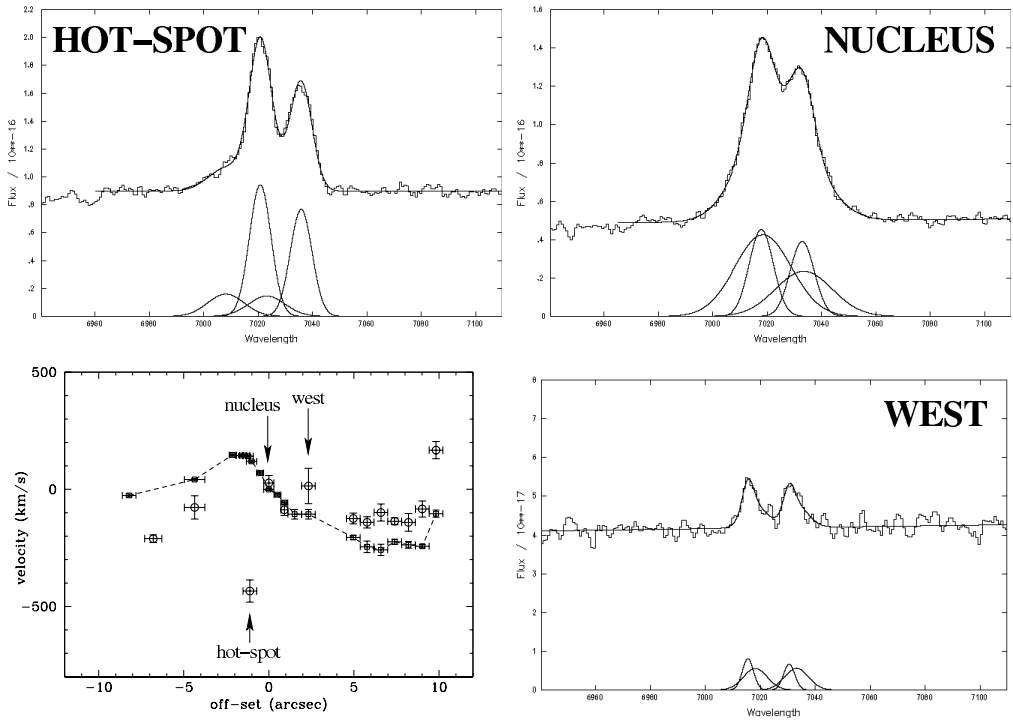}
\caption{Line-fitting results of the [SII]-line in the spectrum along the inner radio axis (p.a. 93$^{\circ}$). 2-component fits to the [SII]-doublet are shown for the bright radio hot-spot, the nucleus and west of the nucleus. {\em Bottom left:} The central velocity (w.r.t. v$_{sys}$) of both the narrow component (small squares connected with the dotted line) and the additional component (circles) plotted against the off-set from the nucleus.}
\label{fig:rotcurve93}
\end{figure}

The outflow is also seen in the [SII]$_{\lambda \lambda 6716+6732}$ doublet. Along the  inner radio axis (p.a. 93$^{\circ}$) the position of the different spectral features is known quite accurately by comparing the shape of the continuum in the optical spectra around the [SII]-line with the the HST image (zoom in of Fig.\ref{fig:ionized}). Like for [OII], the outflow in [SII] is clearly located at the position of the radio hot-spot, $\sim$1 kpc east of the nucleus, as can be seen in Figure \ref{fig:rotcurve93}. Also the narrow component is seen in [SII] along the 93$^{\circ}$ spectrum, showing the kinematics of rotation (projected onto this axis). In addition, on the eastern side (further away from the nucleus) a more modest outflow of gas is seen. On the entire western side of the nucleus the emission-line is much fainter, but it is asymmetric with a redshifted wing. This could either indicate a very mild outflow of order 100 km/s (in this case possibly due to a receding radio jet) or it could be just the effect of integration along the line of sight. At the position of the nucleus the emission line profile is significantly broadened, but very symmetric without any sign of outflow. This might merely reflect a very dynamic state of the ionized gas around the nucleus \citep[a similar line-broadening is also seen in the \HI\ absorption lines on top of the nuclear radio emission,][]{bes02}. In a forthcoming paper \citep{emo05} we will give a full kinematical analysis of the ionized gas in 3C 293. 

\section{Neutral Hydrogen Outflow}
\label{sec:hydrogen}

In Paper 1 we describe the detection of an outflow of neutral hydrogen gas from the central region of 3C 293. This outflow has been detected with the 20MHz band of the WSRT as very broad but very shallow \HI\ absorption against the strong central radio continuum. 
\begin{figure}
\plotone{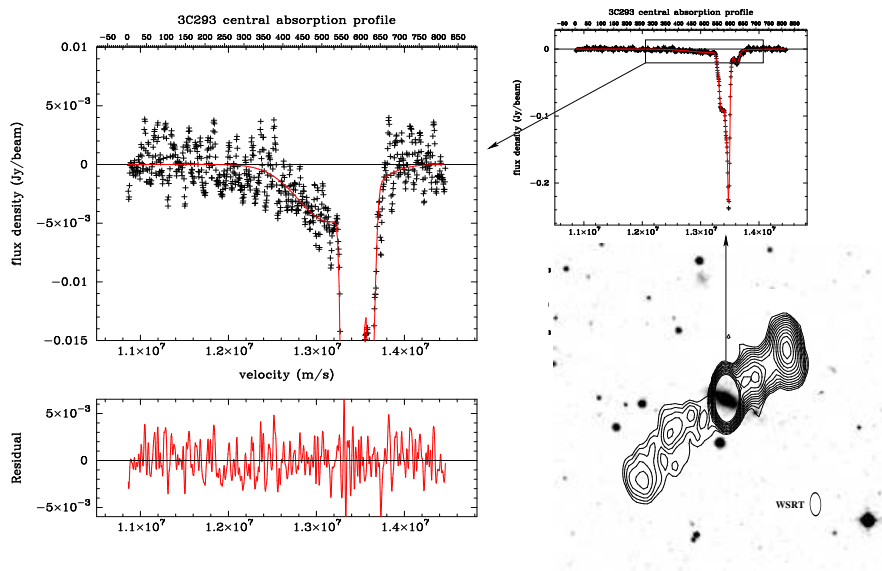}
\caption{Fit to the \HI\ absorption profile against the central region of 3C 293. The left plot shows a zoom-in on the broad component. The line (clearly visible in red in the electronic version of this paper) represents the best fit to the profile. The residuals to the fit (bottom left) consist only of noise.}
\label{fig:profilefit3}
\end{figure}
In Figure \ref{fig:profilefit3} we show a 6-component Gaussian fit to the \HI\ absorption profile from 1 hour of data taken in January 2003. The narrow, deep components have already been studied in detail by \citet{bes04,bes02} and \citet{has85} with higher spatial resolution. This enabled them to spatially disentangle these (plus several other) components, allowing a much more accurate fit to each component separately than we can achieve with our WSRT data. For the broad outflow component our fit gives a FWHM of 852 $\pm$ 41 km/s and central velocity of 13,156 $\pm$ 24 km/s, therefore blue-shifted by 345 km/s w.r.t. the deepest \HI\ absorption. Due to the limited spatial resolution of the WSRT we are not able to determine the exact location of this \HI\ outflow. Nevertheless, the \HI\ outflow is remarkably similar to the outflow of ionized gas at the position of the radio hot-spot (see also Fig.2 of Paper 1). This suggests that one and the same mechanism is likely responsible for the outflow of both the ionized as well as the neutral hydrogen gas.

\section{Companions in \HI\ Emission}
\label{sec:environment}

\begin{figure}[t]
\plotone{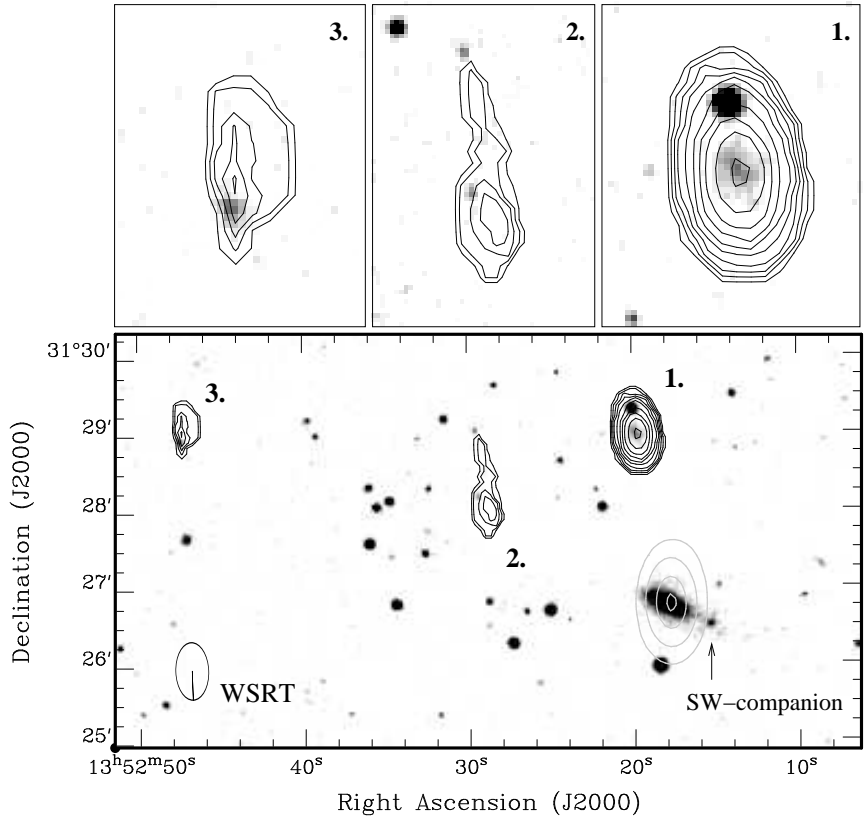}
\caption{Total intensity map of the \HI\ (contours) overlaid on an optical DSS image (gray-scale). Black contours are \HI\ emission; gray contours indicate \HI\ absorption against the radio continuum (shown in Fig.\ref{fig:ionized}). Contour levels are 0.02, 0.035, 0.05, 0.065, 0.08, 0.10, 0.13, 0.16, 0.19, 0.22 and -1, -5, -15, -25 mJy/beam$\times$km/s.} 
\label{fig:fig_emission1}
\end{figure}
Using older 10 MHz WSRT observations (May 7th 2001) we detect three nearby companions of 3C 293 in \HI. Figure \ref{fig:fig_emission1} (top) shows a total intensity plot of the \HI\ gas in these companions. Optical counterparts for companions 1 and 3 are listed in the literature \citep{pat00}. Also companion 2 appears to have an optical counterpart in Fig.\ref{fig:fig_emission1}, but this is hard to tell solely from the DSS image. To our knowledge the redshifts of these companions were up to now unknown. Table \ref{tab:companions} gives the RA and DEC, total \HI\ mass content, projected distance to 3C 293 and velocity of the three companions. 

Given the relatively narrow band of these observations (that is therefore mostly filled by the broad \HI\ absorption) the quality of the final data-cube is not ideal for looking for very faint \HI\ emission, in particular not near the strong ($\sim$3.8 Jy) radio continuum nuclear region of 3C 293. New observations will be necessary to investigate the presence of faint \HI\ emission directly around 3C 293 and its close south-western companion (VV 5-33-12), toward which the optical disk of 3C 293 seems to extend \citep[e.g.][]{eva99,bre84}.

Merger and interaction events are often invoked as the trigger mechanism of AGN activity. Also 3C 293 has often been suggested to be the result of a gas-rich (major) merger event. This is indicated by the complex dust-lanes and the presence of large amounts of dense molecular and neutral gas \citep{eva99,bes02} and by the detection of a young stellar population \citep{tad05}. Gas-rich mergers are likely to produce long-lived signatures in the form of \HI\ tails, bridges or disks (see e.g. the case of the Antennae, Hibbard et al. 2001, or cases of radio galaxies like B2 0648+27, Morganti et al. 2003a,c). Tidal \HI\ tails may even form condensations evolving into (dwarf) galaxies \citep{hib01}. We do not see any tails or other remnant \HI\ feature around 3C 293, at least not at the sensitivity of our present observations. The presence of three \HI\ companions, all located within 400 kpc and 300 km/s, appears similar to e.g. the case of the NGC 3656 group \citep{bal01} where a number of \HI\ companions could be gravitationally bound to the system.
 \begin{table}
\caption{Companions of 3C 293}
\smallskip
\begin{center}
{\small
\begin{tabular}{cccccc}
\tableline
\noalign{\smallskip}
$\#$ & RA & DEC & M$_{HI}$ & Distance & Velocity \\
  & (h:m:s) & ($^{\circ}$:':'')& (M$_{\odot}$) & (kpc) & (km/s)   \\
\noalign{\smallskip}
\tableline
\noalign{\smallskip}
1 & 13:52:20 & +31:29:00 & 1.6$\times$10$^{9}$ & 124 & 13,416 \\
2 & 13:52:29 & +31:28:11 & 3.9$\times$10$^{8}$ & 170 & 13,300 \\
3 & 13:52:47 & +31:28:57 & 2.3$\times$10$^{8}$ & 366 & 13,147 \\
\noalign{\smallskip}
\tableline
\end{tabular}
}
\end{center}
\label{tab:companions}
\end{table}

\section{Conclusions}
\label{sec:conclusions}

In Paper 1 we already concluded that the most likely driving mechanism for the outflow of the neutral hydrogen gas is an interaction between the radio plasma and the ISM. Other explanations are unlikely: An outflow driven by winds from the 1.0 - 2.5 Gyr young stars in the central region of 3C 293 \citep{tad05} would have to be ``fossil'', in which case it is unlikely limited only to the central region where we see it. An AGN radiation-pressure driven outflow seems unlikely given the fact that the AGN in 3C 293 is probably too weak to create the fast outflow. \citep[The ionization in the central region of 3C 293 appears to be very low;][]{emo05}. Also the far-IR luminosity of 3C 293, L$_{60 \mu m}$/L$_{\odot}$ = 10.08, is only at the lower end of the luminosity for quasars; Neugebauer et al. 1986.)

Our analysis of the ionized gas gives additional observational evidence in favor of a jet-ISM interaction as the driving mechanism for the outflow. If the outflowing ionized and neutral gas are indeed different phases of the same physical phenomenon, this shows that despite the high energies that must be involved in the jet-ISM interaction, part of the gas remains, or becomes again, neutral. A possible explanation of this effect is given by \citet{mel02}, who simulate an ionized cloud that is overtaken by a shock-front due to a jet-cloud interaction. Shocks travel into the cloud and fragment it. Due to rapid cooling, dense clumps of gas are formed that that can survive for a long time and that reach velocities up to 500 km/s. The models by \citet{mel02} give credibility to jet-induced star-formation. In the light of our observations it would be nice to further investigate if the dense clumps of gas could consist of neutral hydrogen and if velocities up to $\sim$1000 km/s can be reached.

\vspace{0.2cm}
In a forthcoming paper \citep{emo05} we will give a full analysis of the ionized gas.

\vspace{0.4cm}
\acknowledgments{The author would like to thank J. Holt for her help with the optical spectra, R. Beswick for providing his MERLIN-HST overlay, and J. van Gorkom for a useful discussion on the \HI\ rich companions. BE also acknowledges the University of Sheffield and ASTRON for their hospitality during this project. Part of this research was funded by the Netherlands Organization for Scientific Research (NWO) under grant R 78-379.}

\end{document}